\documentclass[a4paper,12pt]{article}
\usepackage{graphicx,bm, color}
\usepackage{amsmath,amsfonts,amssymb}
\newcommand{\ve}{\bm{\xi}}
\newcommand{\vp}{\bm{\xi'}}
\newcommand{\vs}{\bm{\xi}_{*}}
\newcommand{\vsp}{\bm{\xi'}_{*}}
\newcommand{\bn}{\mathbf{n}}
\newcommand{\bV}{\mathbf{V}}
\newcommand{\bx}{\mathbf{x}}
\newcommand{\bC}{\mathbf{C}}
\newcommand{\bu}{\mathbf{u}}
\newcommand{\bpi}{\bm{\pi}}
\newcommand{\vars}{s}
\newcommand{\rea}{\mathbb{R}}
\newcommand{\nat}{\mathbb{N}}
\newcommand{\irn}[1]{\int_{\rea^{#1}}}
\newcommand{\setve}{D} 
\newcommand{\ivn}[1]{\int_{\setve^{#1}}}
\newcommand{\Ke}{B}
\newcommand{\vol}{\mu}
\newcommand{\p}{\, .}
\newcommand{\sv}{\, ,}
\newcommand{\eq}[1]{(\ref{#1})}
\newcommand{\dm}{\displaystyle}
\newcommand{\ot}{\mbox{$\frac{1}{2}$}}
\newcommand{\sep}{\, |  \,}
\newcommand{\perogni}{\forall \,}
\newcommand{\fr}{\\[5pt]}
\newcommand{\dimostra}{{\sc Proof. }}
\newcounter{npar}
\newcommand{\para}[1]
     {\addtocounter{npar}{1} {\large \bf \thenpar.  #1 } \\}
\newcounter{nprop}
\newenvironment{prop}{
\addtocounter{nprop}{1} {\bf Property \thenprop.} \em}{}
\author{Armando Majorana
\fr Department of Mathematics and Computer Science, \\
University of Catania, Italy}
\title{}
\title{A semi-discrete Boltzmann equation based on a finite volume scheme}
\begin{document}
\baselineskip=20pt
\maketitle
\setcounter{npar}{0}
\setcounter{nprop}{0}
%
\begin{abstract}
In this work we consider the classical non-linear Boltzmann equation, where the unknown is 
the distribution function $f$, which depends on the time $t$, the vector $\bx$ (the 
position of a molecule) and its velocity $\ve$.
From the Boltzmann equation we derive a semi-discrete model, which consists in a set of 
partial differential equations in time and space.
The new unknowns are two moments of the distribution function $f$ and, according to a finite 
volume scheme, integrals of $f$ with respect to the velocity $\ve$, over bounded and open 
sets. 
We do not introduce any approximation for the partial derivatives with respect to $\bx$ and 
the time $t$.

We also propose the use of truncated octahedra as bounded domains of integration.
This reduces both the computing complexity and the number of the constant numerical 
coefficients arising from the collision operator. 
All the numerical para\-meters can be obtained by means of numerical quadrature.
\end{abstract}
MSC-class: 76P, 82C40 (Primary) 65M60 (Secondary)
%
%
%
%
%
\clearpage \noindent
\para{Introduction and basic equations}
We consider the classical non-linear Boltzmann equation (see, for instance \cite{C88}, 
\cite{C90}) that, for neutral monatomic gases, writes
\begin{equation}
\frac{\partial f}{\partial t} + \ve \cdot \frac{\partial f}{\partial \bx} =
Q(f,f)  \p \label{eqb}
\end{equation}
The unknown is the one-particle distribution function $f$, which depends on time $t$, 
position $\bm{x}$ and velocity $\ve$.
The collision operator is defined as follows
\begin{equation}
Q(f,f) = \irn{3} \irn{3} \irn{3}  W(\ve, \vs \sep \vp, \vsp) 
\left( f' f'_{*} - f f_{*} \right)  d \vs \, d \vp \, d \vsp \p
\label{Qff} 
\end{equation}
Here, as in the following, to simplify the notation, often we omit to write the variables $t$
and $\bx$, and sometimes also the velocity $\ve$, explicitly.
As usually, it is here understood that
$$
 f' = f(t,\bx, \vp) , \quad
 f'_{*} = f(t,\bx, \vsp) , \quad
 f = f(t,\bx, \ve) , \quad
 f_{*} = f(t,\bx, \vs) \p
$$
The kernel $W$ of the collision operator is defined by
\begin{equation}
W(\ve, \vs \sep \vp, \vsp) = \Ke(\bn \cdot \bV, |\bV|) \,
\delta(\ve + \vs - \vp - \vsp) \,
\delta(|\ve|^{2} + |\vs|^{2} - |\vp|^{2} - |\vsp|^{2})
 \label{Wv4}  
\end{equation}
where
\begin{equation}
\bn = \dfrac{\ve - \vp}{|\ve - \vp|}  \quad \mbox{and} \quad 
\bV = \ve - \vs \p \label{nV}
\end{equation}
The function $\Ke$ is related to the interaction law between colliding particles. 
In the case of \emph{Variable Hard Sphere} (VHS), $\Ke$ depends only on the modulus of $\bV$ 
and we write $\mathsf{\Ke}(|\bV|)$ instead of $\Ke(\bn \cdot \bV, |\bV|)$.
In particular, if $\dm \mathsf{\Ke}(|\bV|)$ is constant, then we obtain the rigid sphere 
case. 
The two Dirac distributions $\delta$ guarantee the conservation of the momentum and the 
energy during the binary collisions.
The collision operator $Q(f,f)$ of the Boltzmann equation \eq{eqb} is often written in a 
different form, where the Dirac distributions do not appear inside the integral; this 
reduces 
the collision integral to a five-fold integral. 
Here we prefer the form \eq{Qff}, since the symmetric properties are more evident.
Indeed it is easily to prove that
$$
W(\ve, \vs \sep \vp, \vsp) = W(\vp, \vsp \sep \ve, \vs) 
\quad \mbox{and} \quad
W(\ve, \vs \sep \vp, \vsp) = W(\vs, \ve \sep \vsp, \vp) \p
$$

Let $\phi : \rea^{3}  \rightarrow \rea$ be a measurable function. 
Now we assume that the generic test function $\phi$ depends only on the velocity $\ve$, and 
all integrals, involving the distribution function $f$, exist.
If we multiply both sides of the Boltzmann equation \eq{eqb} by a test function $\phi(\ve)$
and we formally integrate over $\rea^{3}$ with respect to the velocity $\ve$, then we obtain 
the following weak form of the Boltzmann equation
\begin{equation}
\frac{\partial \mbox{ }}{\partial t} \irn{3} f(t,\bx, \ve) \, \phi(\ve) \: d \ve 
+ \frac{\partial \mbox{ }}{\partial \bx}
\irn{3}  f(t,\bx, \ve) \, \ve \, \phi(\ve) \: d \ve =
\irn{3} Q(f,f)(t,\bx, \ve) \, \phi(\ve) \: d \ve \p 
\label{eqBve}
\end{equation}
Hence, for any test function $\phi$, we obtain an equation of type \eq{eqBve}.\\
In order to simplify the formulas, we set $\setve = \rea^{3}$, 
$\setve^{2} = \setve \times \rea^{3}$, ...,
$\setve^{n+1} = \setve^{n} \times \rea^{3}$.
It is useful to recall a well-known result.
The right hand side of Eq.~\eq{eqBve} can be written as follows
$$
\ot \ivn{4} W(\ve, \vs | \vp, \vsp)
\left[ \phi(\vp) + \phi(\vsp) - \phi(\ve) - \phi(\vs) \right]
f(t,\bx, \ve) \,  f(t,\bx, \vs) \: d \vp \, d \vsp \, d \vs \, d \ve 
$$
or
\begin{align}
& 
\ot \ivn{4}  W(\ve, \vs | \vp, \vsp)
\left[ \phi(\vp) + \phi(\vsp) \right]
f(t,\bx, \ve) \,  f(t,\bx, \vs) \: d \vp \, d \vsp \, d \vs \, d \ve 
\nonumber
\fr
& \mbox{} -
\ot \ivn{4} W(\ve, \vs | \vp, \vsp)
\left[ \phi(\ve) + \phi(\vs) \right]
f(t,\bx, \ve) \,  f(t,\bx, \vs) \: d \vp \, d \vsp \, d \vs \, d \ve \p
\label{Qphi}
\end{align}
Now, if we define
\begin{equation}
G(\phi; \ve , \vs) = 
\ivn{2} W(\ve, \vs | \vp, \vsp)
\left[ \phi(\vp) + \phi(\vsp) \right] d \vp \, d \vsp \sv
\label{Gphi} 
\end{equation}
and (see Appendix A for the details)
\begin{equation}
\nu(|\bV|) =  
\ivn{2} W(\ve, \vs | \vp, \vsp) \: d \vp \, d \vsp \sv
\label{tot_cs} 
\end{equation}
then integrals \eq{Qphi} write
\begin{align*}
&
\ot \int_{\setve^{2}} G(\phi; \ve , \vs) \, f(t,\bx, \ve) \, f(t,\bx, \vs) \: d \vs \, d \ve 
\fr
&
\mbox{} -
\ot \int_{\setve^{2}} \nu(|\bV|) \left[ \phi(\ve) + \phi(\vs) \right] 
f(t,\bx, \ve) \, f(t,\bx, \vs) \: d \vs \, d \ve \p 
\end{align*}
It is evident that the first integral corresponds to the gain term of the collision 
operator, and the second to the loss term.
Now Eq.~\eq{eqBve} becomes
\begin{align}
&
\frac{\partial \mbox{ }}{\partial t} \ivn{} f(t,\bx, \ve) \, \phi(\ve) \: d \ve 
 + \frac{\partial \mbox{ }}{\partial \bx} \! \ivn{} 
f(t,\bx, \ve) \, \ve \, \phi(\ve) \: d \ve
\nonumber
\fr
& \hspace{30pt} = 
\ot \ivn{2} G(\phi; \ve , \vs) \, f(t,\bx, \ve) \, f(t,\bx, \vs) \: d \vs \, d \ve 
\nonumber
\fr
& \hspace{30pt} -
\ot \ivn{2} \nu(|\bV|) \left[ \phi(\ve) + \phi(\vs) \right] 
f(t,\bx, \ve) \, f(t,\bx, \vs) \: d \vs \, d \ve \p 
\label{BTEphi}
\end{align}
We note that the function $G(\phi; \ve , \vs)$, which is strictly related to the function
$\phi$, depends only on the $(\ve, \vs)$ variables, and plays the role of the kernel of the
integral operator corresponding to the gain term of the collision operator. \\
We recall a simple but important result.
If we denote by $\psi(\ve)$ any of the five collision invariants
$\dm \left\{ 1, \ve, | \ve |^{2} \right\} $, 
then it is easily to prove that
\begin{equation}
G(\psi; \ve , \vs) = \nu(|\bV|) \left[ \psi(\ve) + \psi(\vs) \right] 
\quad \forall \, \ve , \vs \in \setve \p 
\label{Gnu} 
\end{equation}
%
%
%
%
%
%
%
\para{The semi-discrete model}
We consider a partition of the three-dimensional real space $\setve$.
Let $\left\lbrace C_{\alpha} \right\rbrace$ be a {\bf numerable} family of open, not empty,
bounded subsets such that
$$
C_{\alpha} \subseteq \setve \quad \perogni \alpha \in \nat \sv \quad
C_{\alpha} \cap C_{\beta} = \emptyset 
\quad \perogni \alpha \neq \beta \quad \mbox{and }
\bigcup_{\alpha = 1}^{\infty} \overline{C_{\alpha}} = \setve \p
$$
Moreover, we assume that 
\begin{equation}
\mbox{
\emph{
for every compact $K \subseteq\setve$, 
there exists an integer $n_{K}$ such that
$ \dm K \subseteq \bigcup_{\alpha = 1}^{n_{K}} \overline{C_{\alpha}}$. 
}}
\label{h1}
\end{equation}
We denote by $\chi_{\alpha}$ the characteristic function of the cell $C_{\alpha}$, that is
$$
\chi_{\alpha}(\ve) = \left\{
\begin{array}{ll}
1 & \mbox{ if } \ve \in C_{\alpha}
\\[5pt]
0 & \mbox{ otherwise}
\end{array}
\right.
$$
and we choose the following test functions
$$
\chi_{\alpha}(\ve) \: (\mbox{for } \alpha = 1,2,3,... ) , \:
\ve , \mbox{ and } \ot |\ve|^{2}  \p
$$
A similar approach was also proposed in \cite{AM, AMa}. 
Therefore Eq.~\eq{BTEphi} gives the system
\begin{align}
&
\dfrac{\partial \mbox{ }}{\partial t} \int_{C_{\alpha}} f(t, \bx, \ve) \: d \ve 
+ \dfrac{\partial \mbox{ }}{\partial \bx} \cdot 
\int_{C_{\alpha}} f(t, \bx, \ve) \, \ve \: d \ve \nonumber
\fr
&
\hspace{10pt} = \ot \ivn{2} 
G(\chi_{\alpha}; \ve , \vs) \, f(t, \bx, \ve) \, f(t, \bx, \vs) \: d \vs \, d \ve
\nonumber
\fr
& 
\hspace{10pt} -
\ot \ivn{2} \nu(|\bV|) \left[ \chi_{\alpha}(\ve) + \chi_{\alpha}(\vs) \right] 
f(t, \bx, \ve) \, f(t, \bx, \vs) \: d \vs \, d \ve \qquad (\alpha = 1,2,3,... )
\label{eqa}
\fr
&
\dfrac{\partial \mbox{ }}{\partial t} \ivn{} f(t, \bx, \ve) \, \xi_{i} \: d \ve 
+ \dfrac{\partial \mbox{ }}{\partial \bx} \cdot 
\ivn{} f(t, \bx, \ve) \, \ve \, \xi_{i} \: d \ve = 0 \hspace{67pt} (i = 1,2,3)
\label{mom}
\fr
&
\dfrac{\partial \mbox{ }}{\partial t} \ivn{} f(t, \bx, \ve) \, \ot |\ve|^{2} \: d \ve 
+ \dfrac{\partial \mbox{ }}{\partial \bx} \cdot 
\ivn{} f(t, \bx, \ve) \, \ve \, \ot |\ve|^{2} \: d \ve = 0 \p
\label{dense}
\end{align}
We can simplify Eq.~\eq{eqa}, since
\begin{align*}
&
\ivn{2} \nu(|\bV|) \left[ \chi_{\alpha}(\ve) + \chi_{\alpha}(\vs) \right] 
f(t, \bx, \ve) \, f(t, \bx, \vs) \: d \vs \, d \ve 
\fr
& \mbox{} = \!
\ivn{2} \! \nu(|\bV|) \, \chi_{\alpha}(\ve) \, 
f(t, \bx, \ve) \, f(t, \bx, \vs) \, d \vs \, d \ve + \! \!
\ivn{2} \! \nu(|\bV|) \, \chi_{\alpha}(\vs) \,
f(t, \bx, \ve) \, f(t, \bx, \vs) \, d \vs \, d \ve
\fr
& \mbox{} =
\int_{C_{\alpha}} \ivn{} \nu(|\bV|) \, f(t, \bx, \ve) \, f(t, \bx, \vs) \: d \vs \, d \ve +
\ivn{} \int_{C_{\alpha}} \nu(|\bV|) \, f(t, \bx, \ve) \, f(t, \bx, \vs) \: d \vs \, d \ve
\fr
& \mbox{} = 2 \int_{C_{\alpha}} \ivn{}  \nu(|\bV|) \, 
f(t, \bx, \ve) \, f(t, \bx, \vs) \: d \vs \, d \ve \p
\end{align*}
Now we define the following set of functions (integrals of the distribution function $f$)
\begin{align}
&
N_{\alpha}(t, \bx) :=  \int_{C_{\alpha}} f(t, \bx, \ve) \: d \ve  \hspace{24pt}
(\alpha = 1,2,3,... ) \sv \label{Na}
\fr
&
p_{i}(t, \bx) := \ivn{} f(t, \bx, \ve) \, \xi_{i} \: d \ve \qquad (i = 1,2,3) \sv
\label{pi}
\fr
&
{\cal{E}}(t, \bx) := \ot \ivn{} f(t, \bx, \ve) \, |\ve|^{2} \: d \ve \p
\label{En}
\end{align}
The physical meaning of $N_{\alpha}(t, \bx)$ is evident: it is the density of the particles, 
that have velocity $\ve$ belonging to the cell $C_{\alpha}$ at time $t$ and position $\bx$. 
The variables $p_{i}(t, \bx)$ and ${\cal{E}}(t, \bx)$ are two classical moments of the 
distribution function $f$. \\
The aim of this work is to derive a set of partial differential equations for the new 
unknowns $N_{\alpha}$, $p_{i}$ and ${\cal{E}}$.
Introducing the variables \eq{Na}-\eq{En}, then the set of equations \eq{eqa}-\eq{dense} 
becomes
\begin{align}
&
\dfrac{\partial N_{\alpha}}{\partial t} \, (t, \bx)
+ \dfrac{\partial \mbox{ }}{\partial \bx} \cdot 
\int_{C_{\alpha}} f(t, \bx, \ve) \, \ve \: d \ve = \ot \ivn{2}
G(\chi_{\alpha}; \ve , \vs) \, f(t, \bx, \ve) \, f(t, \bx, \vs) \: d \vs \, d \ve
\nonumber
\fr
& 
\hspace{10pt} -
\int_{C_{\alpha}} \ivn{}  \nu(|\bV|) \, 
f(t, \bx, \ve) \, f(t, \bx, \vs) \: d \vs \, d \ve
\quad (\alpha = 1,2,3,... )
\label{eqNf2}
\fr
&
\dfrac{\partial p_{i}}{\partial t} \, (t, \bx) 
+ \dfrac{\partial \mbox{ }}{\partial \bx} \cdot 
\ivn{} f(t, \bx, \ve) \, \ve \, \xi_{i} \: d \ve = 0 
\hspace{35pt} (i = 1,2,3)
\label{eqpf2}
\fr
&
\dfrac{\partial {\cal{E}}}{\partial t} \, (t, \bx) 
+  \ot \, \dfrac{\partial \mbox{ }}{\partial \bx} \cdot 
\ivn{} f(t, \bx, \ve) \, \ve \, |\ve|^{2} \: d \ve = 0 \p
\label{eqEf2}
\end{align}
Of course, the system of Eqs.~\eq{eqNf2}-\eq{eqEf2} is not {\bf closed}. \\
According to the finite volume scheme, the simplest way to derive a closed system is 
to introduce the following set of numerical parameters, \emph{which do not depend on the 
distribution function} $f$,
\begin{align}
G_{\alpha; \beta, \gamma} & =
\dfrac{1}{\vol(C_{\beta})} \, \dfrac{1}{\vol(C_{\gamma})}
\int_{C_{\beta}} \int_{C_{\gamma}} G(\chi_{\alpha}; \ve , \vs) \: d \vs \, d \ve \sv 
\quad (\alpha, \beta, \gamma \in \nat)
\fr
\nu_{\alpha; \beta} & =
\dfrac{1}{\vol(C_{\alpha})} \, \dfrac{1}{\vol(C_{\beta})}
\int_{C_{\alpha}} \int_{C_{\beta}} \nu(|\bV|) \: d \vs \, d \ve \sv 
\hspace{37pt} (\alpha, \beta \in \nat)
\fr
\bm{m}_{\alpha} & =
\dfrac{1}{\vol(C_{\alpha})} \int_{C_{\alpha}} \! \ve \: d \ve , 
\hspace{141pt} (\alpha \in \nat)
\fr
\bpi_{\alpha, i} & =
\dfrac{1}{\vol(C_{\alpha})} \int_{C_{\alpha}} \! \ve \, \xi_{i} \: d \ve ,  
\hspace{130pt} (\alpha \in \nat \sv i = 1,2,3)
\fr
\bm{e}_{\alpha} & =
\dfrac{1}{\vol(C_{\alpha})} \int_{C_{\alpha}} \! \ve \, |\ve|^{2} \: d \ve , 
\hspace{122pt} (\alpha \in \nat)
\end{align}
where $\vol(C_{\alpha})$ is the measure of the cell $C_{\alpha}$. \\
The parameter $G_{\alpha; \beta, \gamma}$ is an \emph{average} of the function 
$G(\chi_{\alpha}; \ve , \vs)$ over the set
$$
S_{\beta \gamma} =
\left\lbrace 
( \ve, \vs ) \in \setve^{2} : \ve \in C_{\beta} \mbox{ e } \vs \in C_{\gamma} 
\right\rbrace .
$$
The meaning of the other parameters is analogous.
Now we assume reasonable that
\begin{align*}
&
\ivn{2}
G(\chi_{\alpha}; \ve , \vs) \, f(t, \bx, \ve) \, f(t, \bx, \vs) \: d \vs \, d \ve
\\
&
\hspace{15pt} =
\sum_{\beta=1}^{\infty} \sum_{\gamma=1}^{\infty}
\int_{C_{\beta}} \int_{C_{\gamma}} G(\chi_{\alpha}; \ve , \vs) \, 
f(t, \bx, \ve) \, f(t, \bx, \vs) \: d \vs \, d \ve 
\\
&
\hspace{15pt} \approx
\sum_{\beta=1}^{\infty} \sum_{\gamma=1}^{\infty}
\int_{C_{\beta}} \int_{C_{\gamma}} G_{\alpha; \beta, \gamma} \,
f(t, \bx, \ve) \, f(t, \bx, \vs) \: d \vs \, d \ve  
=
\sum_{\beta=1}^{\infty} \sum_{\gamma=1}^{\infty}
G_{\alpha; \beta, \gamma} \, N_{\beta}(t, \bx) \, N_{\gamma}(t, \bx) \sv
\fr
&
\int_{C_{\alpha}} \ivn{}  \nu(|\bV|) \, 
f(t, \bx, \ve) \, f(t, \bx, \vs) \: d \vs \, d \ve
 =
\sum_{\beta=1}^{\infty} \int_{C_{\alpha}} \int_{C_{\beta}}  \nu(|\bV|) \, 
f(t, \bx, \ve) \, f(t, \bx, \vs) \: d \vs \, d \ve
\\
&
\hspace{15pt} \approx
\sum_{\beta=1}^{\infty} \int_{C_{\alpha}} \int_{C_{\beta}} \nu_{\alpha; \beta} \, 
f(t, \bx, \ve) \, f(t, \bx, \vs) \: d \vs \, d \ve
 = 
\sum_{\beta=1}^{\infty} \nu_{\alpha; \beta} \, N_{\alpha}(t, \bx) \,  N_{\beta}(t, \bx) ,
\fr
&
\int_{C_{\alpha}} f(t, \bx, \ve) \, \ve \: d \ve \approx
\bm{m}_{\alpha} \, N_{\alpha}(t, \bx) ,
\fr
&
\ivn{} f(t, \bx, \ve) \, \ve \, \xi_{i} \: d \ve =
\sum_{\alpha=1}^{\infty} \int_{C_{\alpha}} f(t, \bx, \ve) \, \ve \, \xi_{i} \: d \ve
\approx \sum_{\alpha=1}^{\infty} \bpi_{\alpha, i} \, N_{\alpha}(t, \bx) ,
\fr
&
\ivn{} f(t, \bx, \ve) \, \ve \, |\ve|^{2} \: d \ve =
\sum_{\alpha=1}^{\infty} \int_{C_{\alpha}} f(t, \bx, \ve) \, \ve \, |\ve|^{2} \: d \ve
\approx \sum_{\alpha=1}^{\infty} \bm{e}_{\alpha} \, N_{\alpha}(t, \bx) .
\end{align*}
Introducing these approximations, the system \eq{eqNf2}-\eq{eqEf2} gives the {\bf closed} 
system of partial differential equations
\begin{align}
&
\frac{\partial N_{\alpha}}{\partial t} (t, \bx)  + 
\bm{m}_{\alpha} \cdot \frac{\partial N_{\alpha}}{\partial \bx} (t, \bx)
\nonumber
\fr
& \hspace{10pt}
= \ot \sum_{\beta=1}^{\infty} \sum_{\gamma=1}^{\infty}
G_{\alpha; \beta, \gamma} \, N_{\beta}(t, \bx) \, N_{\gamma}(t, \bx)  -  
\sum_{\beta=1}^{\infty} \nu_{\alpha; \beta} \, N_{\alpha}(t, \bx) \, N_{\beta}(t, \bx) 
\quad (\alpha = 1, 2, , ... )
\label{eqNfa}
\fr
&
\dfrac{\partial p_{i}}{\partial t} (t, \bx) 
+ \sum_{\alpha=1}^{\infty} \bpi_{\alpha, i} \cdot
\dfrac{\partial N_{\alpha}}{\partial \bx} (t, \bx) = 0 
\hspace{161pt} (i = 1,2,3)
\label{eqpfa}
\\[12pt]
&
\dfrac{\partial {\cal{E}}}{\partial t} (t, \bx) 
+ \ot \sum_{\alpha=1}^{\infty} \bm{e}_{\alpha}\cdot
\dfrac{\partial N_{\alpha}}{\partial \bx} (t, \bx) = 0 \p
\label{eqEfa}
\end{align}
We remark that Eqs.~\eq{eqNfa} are Smoluchowski type equations; they do not contain the 
other unknowns $p_{i}$ and $\cal{E}$. So, the system of equations \eq{eqNfa}-\eq{eqEfa} is 
decoupled. \fr
%
%
%
%
%
%
%
\para{Conservation laws}
The model \eq{eqNfa}-\eq{eqEfa} retains some of the main features of the Boltzmann 
equation. This is a consequence of the fact that the numerical parameters 
$G_{\alpha; \beta, \gamma}$ and $\nu_{\alpha; \beta}$ are related to the collision operator 
$Q(f,f)$.
Firstly we observe that Eqs.~\eq{eqpfa}-\eq{eqEfa} correspond to the conservation of 
momentum and energy, that are constant for spatially homogeneous solutions.
Also the conservation of the mass is recovered at the discrete level.
This is shown after proving two simple results. \\
\begin{prop}
For every pair $(\beta, \gamma)$, the parameters $G_{\alpha; \beta, \gamma}$ are zero except 
at most a finite number of indexes $\alpha$.
\end{prop}
\\
\dimostra
In the Appendix A, we show that
\begin{align*}
G(\chi_{\alpha}; \ve , \vs) & = 
\dfrac{|\bV|}{8} \int_{-1}^{1} d \vars \int_{-\pi}^{\pi} d \theta \:
\Ke(\ot \left| \bV - |\bV| \, \bu \right|, |\bV|)
\fr
& \times
\left[ \chi_{\alpha} \! \left( \ot (\ve + \vs) + \ot |\bV| \, \bu \right) + 
\chi_{\alpha} \! \left( \ot (\ve + \vs) - \ot |\bV| \, \bu \right) \right] ,
\end{align*}
where 
$\bu= \left( \sqrt{1 - \vars^{2}} \, \cos \theta , \sqrt{1 - \vars^{2}} \, \sin \theta , 
\vars \right)$ 
is an unit vector.
We choose the two indexes $\beta$ e $\gamma$ arbitrarily, and we consider the restriction of 
the function $G(\chi_{\alpha}; \ve , \vs)$ to the set
$$
S_{\beta \gamma} =
\left\lbrace 
( \ve, \vs ) \in \setve^{2} : \ve \in C_{\beta} \mbox{ e } \vs \in C_{\gamma} 
\right\rbrace .
$$
Since the cells are bounded sets, there exists a constant $c_{\beta \gamma}$, such that
\begin{center}
$ \dm |\ve|^{2} + |\vs|^{2} \leq c_{\beta \gamma} $ for every 
$(\ve , \vs) \in S_{\beta \gamma}$. 
\end{center}
Now
\begin{align*}
&
\left| \ot (\ve + \vs) + \ot |\bV| \, \bu \right|^{2} +
\left| \ot (\ve + \vs) - \ot |\bV| \, \bu \right|^{2}
\fr
& =
\ot \left| \ve + \vs \right|^{2} + \ot \left| \bV \right|^{2} =
|\ve|^{2} + |\vs|^{2} \leq c_{\beta \gamma}
\quad \perogni (\ve, \vs) \in S_{\beta \gamma} \mbox{ and } \perogni \bu \in S^{2}.
\end{align*}
Therefore there exists a compact set $K_{\beta \gamma} \subset \setve$, depending only on the 
indexes $\beta$ e $\gamma$, such that
$$
\ot (\ve + \vs) + \ot |\bV| \, \bu \in K_{\beta \gamma} 
\mbox{ and }
\ot (\ve + \vs) - \ot |\bV| \, \bu \in K_{\beta \gamma} ,
\quad \perogni (\ve, \vs) \in S_{\beta \gamma} \mbox{ and } \perogni \bu \in S^{2} .
$$
Hypothesis \eq{h1} guarantees the existence of a finite covering of $K_{\beta \gamma}$ by 
cells; therefore we can find an 
index $\alpha_{\beta \gamma}$ such that
$$
K_{\beta \gamma} \cap C_{\alpha} = \emptyset \quad \perogni \alpha > \alpha_{\beta \gamma} ,
$$
which implies $G_{\alpha; \beta, \gamma} = 0$ for every $\alpha > \alpha_{\beta \gamma}$.
$\blacktriangle$
\fr
\begin{prop}
$$ 
\ot \sum_{\alpha=1}^{\infty} G_{\alpha; \beta, \gamma} = \nu_{\beta; \gamma} 
\qquad \perogni \beta , \gamma \in \nat .
$$
\end{prop}
\dimostra
Due to the Property 1, the series reduces to a finite sum, and this simplifies the proof.
We must prove that
$$
\ot \sum_{\alpha=1}^{\infty} 
\int_{C_{\beta}} \int_{C_{\gamma}} G(\chi_{\alpha}; \ve , \vs) \: d \vs \, d \ve
=
\int_{C_{\beta}} \int_{C_{\gamma}} \nu(|\bV|) \: d \vs \, d \ve
\qquad \perogni \beta , \gamma \in \nat ,
$$
or
$$
\int_{C_{\beta}} \int_{C_{\gamma}} \left[
\sum_{\alpha=1}^{\infty} 
G(\chi_{\alpha}; \ve , \vs) - 2 \, \nu(|\bV|) \right] d \vs \, d \ve = 0
\qquad \perogni \beta , \gamma \in \nat .
$$
Now, using the definition \eq{Gphi}, we have for every $\ve , \vs \in \setve$
\begin{align*}
\sum_{\alpha=1}^{\infty} G(\chi_{\alpha}; \ve , \vs) 
& =
 \sum_{\alpha=1}^{\infty} 
\ivn{2} W(\ve, \vs | \vp, \vsp)
\left[ \chi_{\alpha}(\vp) + \chi_{\alpha}(\vsp) \right] d \vp \, d \vsp
\fr
& =
\sum_{\alpha=1}^{\infty} 
\ivn{} \int_{C_{\alpha}} \! \! W(\ve, \vs | \vp, \vsp) \: d \vp \, d \vsp
+
\sum_{\alpha=1}^{\infty}
\int_{C_{\alpha}} \ivn{} \! \! W(\ve, \vs | \vp, \vsp) \: d \vp \, d \vsp
\fr
& =
2 \, \ivn{2} \! \! W(\ve, \vs | \vp, \vsp) \: d \vp \, d \vsp 
=
2 \, \nu(|\bV|) \p \quad \blacktriangle
\end{align*}
This property is equivalent to Eq.~\eq{Gnu} for $\psi = 1$, and it guarantees the 
conservation of the mass for the semi-discrete model.
Infact from Eq.~\eq{eqNfa} it follows
\begin{align*}
&
\sum_{\alpha=1}^{\infty} \left[ \frac{\partial N_{\alpha}}{\partial t} (t, \bx)  + 
\bm{m}_{\alpha} \cdot \frac{\partial N_{\alpha}}{\partial \bx} (t, \bx) \right]
\fr
& \hspace{10pt}
= \ot \sum_{\alpha=1}^{\infty} \sum_{\beta=1}^{\infty} \sum_{\gamma=1}^{\infty}
G_{\alpha; \beta, \gamma} \, N_{\beta}(t, \bx) \, N_{\gamma}(t, \bx) - 
\sum_{\alpha=1}^{\infty} \sum_{\beta=1}^{\infty} \nu_{\alpha; \beta} \, 
N_{\alpha}(t, \bx) \, N_{\beta}(t, \bx)
\fr
& \hspace{10pt}
= \ot \sum_{\alpha=1}^{\infty} \sum_{\beta=1}^{\infty} \sum_{\gamma=1}^{\infty}
G_{\alpha; \beta, \gamma} \, N_{\beta}(t, \bx) \, N_{\gamma}(t, \bx) - 
\sum_{\beta=1}^{\infty} \sum_{\gamma=1}^{\infty} \nu_{\beta; \gamma} \, 
N_{\beta}(t, \bx) \, N_{\gamma}(t, \bx)
\fr
& \hspace{10pt}
= \sum_{\beta=1}^{\infty} \sum_{\gamma=1}^{\infty} \left[  
\ot \sum_{\alpha=1}^{\infty} G_{\alpha; \beta, \gamma} - \nu_{\beta; \gamma} \right] 
N_{\beta}(t, \bx) \, N_{\gamma}(t, \bx) = 0 \p
\end{align*}

Equations \eq{eqpfa}-\eq{eqEfa} approximate the exact equations, which are obtained
by Eq.~\eq{eqBve} for $\phi(\ve) = \xi_{i}$ $(i=1,2,3)$ and $\phi(\ve) = | \ve |^{2}$.
So the quantities $p_{i}$ and ${\cal{E}}$ are not the \emph{exact} moments of the 
distribution functions, except for spatially homogeneous solutions.
Nevertheless this represents an advantage. 
In fact, let us consider a solution of Eqs.~\eq{eqNfa}-\eq{eqEfa}.
Since, for $i = 1,2,3$,
\begin{align*}
p_{i}(t, \bx) = & \ivn{} f(t, \bx, \ve) \, \xi_{i} \: d \ve =
\sum_{\alpha=1}^{\infty} \int_{C_{\alpha}} f(t, \bx, \ve) \, \xi_{i} \: d \ve
\approx \sum_{\alpha=1}^{\infty} \left[ \dfrac{1}{\mu(C_{\alpha})} 
\int_{C_{\alpha}} \xi_{i} \: d \ve \right] N_{\alpha}(t, \bx) \sv
\end{align*}
then the function 
$$
\left| p_{i}(t, \bx) - \sum_{\alpha=1}^{\infty} \left[ \dfrac{1}{\mu(C_{\alpha})} 
\int_{C_{\alpha}} \xi_{i} \: d \ve \right] N_{\alpha}(t, \bx) \right|
$$
furnishes a simple measure of the goodness of the solution of the model.
Analogously we can use the relationship
\begin{align*}
{\cal{E}}(t, \bx) = & \, \ot \ivn{} f(t, \bx, \ve) \, |\ve|^{2} \: d \ve =
\ot \sum_{\alpha=1}^{\infty} \int_{C_{\alpha}} f(t, \bx, \ve) \, |\ve|^{2} \: d \ve 
\fr
\approx & \, \ot \sum_{\alpha=1}^{\infty} \left[ \dfrac{1}{\mu(C_{\alpha})} 
\int_{C_{\alpha}} |\ve|^{2} \: d \ve \right] N_{\alpha}(t, \bx) \p 
\end{align*}
\clearpage \noindent
%
%
%
%
\para{A domain decomposition}
We propose a regular decomposition, where the cells are truncated octahedra.
\vspace{-77pt} 
\begin{center}
\includegraphics[width=0.8\textwidth]{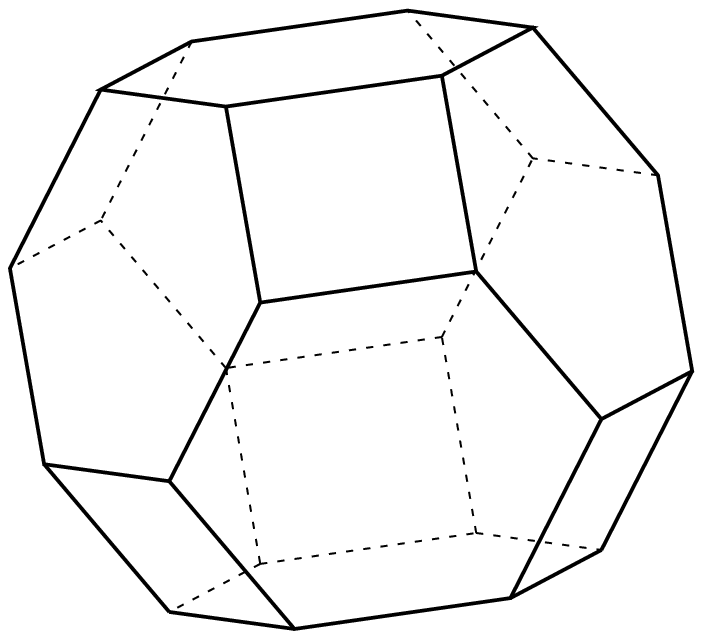}
\\[-77pt]
Figure 1. The truncated octahedron.
\end{center}
The truncated octahedron is the Archimedean solid, with 8 regular hexagonal faces 
and 6 square faces.
Let $\ell$ be the distance between two opposite square faces.
If we denote by $(c_{1}, c_{2}, c_{3})$ the coordinates of the center of a generic truncated 
octahedron, and all the square faces are parallel to the coordinate planes, then 
$(\xi_{1}, \xi_{2}, \xi_{3})$ belongs to the open cell if and only if the following 
inequalities are 
satisfied
\begin{align*}
&
| \xi_{1} - c_{1} | < \dfrac{\ell}{2} \sv \quad
| \xi_{2} - c_{2} | < \dfrac{\ell}{2} \sv \quad
| \xi_{3} - c_{3} | < \dfrac{\ell}{2} \sv 
\\[3pt]
&
| \xi_{1} - c_{1} | + | \xi_{2} - c_{2} | + | \xi_{3} - c_{3} | < 
\dfrac{3}{4} \, \ell \p
\end{align*}
It is fundamental to point out that the truncated octahedron is a space-filling polyhedron.
The number of the nearest cells to every cell is fourteen, whereas, for instance, for cubic 
cells this number is twenty-six.
The coordinates of the centers of the fourteen neighboring cells are 
\begin{align*}
&
\left( c_{1} - \dfrac{1}{2} \, \ell, c_{2} + \dfrac{1}{2} \, \ell, c_{3} \pm \dfrac{\ell}{2} 
\right) , \quad
\left( c_{1} - \dfrac{1}{2} \, \ell, c_{2} - \dfrac{1}{2} \, \ell, c_{3} \pm \dfrac{\ell}{2} 
\right) ,
\\[7pt]
&
\left( c_{1} + \dfrac{1}{2} \, \ell, c_{2} + \dfrac{1}{2} \, \ell, c_{3} \pm \dfrac{\ell}{2} 
\right) , \quad
\left( c_{1} + \dfrac{1}{2} \, \ell, c_{2} - \dfrac{1}{2} \, \ell, c_{3} \pm \dfrac{\ell}{2} 
\right) ,
\\[7pt]
&
\left( c_{1} \pm \ell, c_{2}, c_{3} \right) \sv \quad
\left( c_{1}, c_{2} \pm \ell, c_{3} \right) \sv \quad
\left( c_{1}, c_{2}, c_{3} \pm \ell \right) \p
\end{align*}
Let us denote by $\mathbf{c}_{\beta}$ the center of the cell $C_{\beta}$.
The truncated octahedron has some interesting properties. One of these is the following:
\begin{center}
if $\ve \in C_{\alpha}$, then 
$| \ve - \mathbf{c}_{\alpha}| < | \ve - \mathbf{c}_{\beta} | \quad \perogni \beta \neq 
\alpha$. 
\end{center}
This allows us to easily determine the position of a velocity within a cell. \fr
%
%
%
%
%
%
%
%
%
%
%
%
%
{\large \bf Appendix A} \fr
We recall some well-known formulas, which are useful to transform the collision operator. 
In particular they are used to reduce the collision operator to a five-fold integral.
Let $F$ be a continuous function defined on $\setve^{4}$. We have
\begin{align}
&
\ivn{2} F(\ve, \vs, \vp, \vsp) \, \delta(\ve + \vs - \vp - \vsp) \,
\delta(|\ve|^{2} + |\vs|^{2} - |\vp|^{2} - |\vsp|^{2}) \: d \vp \, d \vsp
\nonumber
\fr
&
\hspace{40pt}
[ \emph{ by means of the change of variables: } \vp = \ve - \bm{A}, \vsp = \vs + \bm{B} ]
\nonumber
\fr
&
= \! \ivn{2} F(\ve, \vs, \ve - \bm{A}, \vs + \bm{B}) \, \delta(\bm{A} - \bm{B}) \,
\delta(|\ve|^{2} + |\vs|^{2} - |\ve - \bm{A}|^{2} - |\vs + \bm{B}|^{2}) 
\: d \bm{A} \, d \bm{B}
\nonumber
\fr
&
= \ivn{} F(\ve, \vs, \ve - \bm{A}, \vs + \bm{A}) \,
\delta(|\ve|^{2} + |\vs|^{2} - |\ve - \bm{A}|^{2} - |\vs + \bm{A}|^{2}) \: d \bm{A}
\nonumber
\fr
&
= \ivn{} F(\ve, \vs, \ve - \bm{A}, \vs + \bm{A}) \,
\delta(-2 \, |\bm{A}|^{2} + 2 \, \bm{A} \cdot \ve - 2 \, \bm{A} \cdot \vs) \: d \bm{A} 
\nonumber
\fr
&
= \ot \ivn{} F(\ve, \vs, \ve - \bm{A}, \vs + \bm{A}) \,
\delta(|\bm{A}|^{2} - \bm{A} \cdot \bV) \: d \bm{A} \tag{A.1} \label{F_A}
\fr
&
\hspace{40pt}
[ \emph{ by using the variable transformation }
\left( \bC = \ot \bV  - \bm{A}\right) ]
\nonumber
\fr
&
= \ot \ivn{} F(\ve, \vs, \ve + \bC - \ot \bV, 
\vs - \bC + \ot \bV) \,
\delta\left( |\bC|^{2} - \mbox{$\frac{1}{4}$} |\bV|^{2} \right) d \bC 
\nonumber
\fr
&
= \ot \ivn{} F(\ve, \vs, \ot (\ve + \vs) + \bC, \ot (\ve + \vs) - \bC ) \,
\delta\left( |\bC|^{2} - \mbox{$\frac{1}{4}$} |\bV|^{2} \right) d \bC 
\tag{A.2} \label{F_C}
\fr
&
\hspace{40pt}
[ \emph{ set } \bC = r \, \bu \emph{ with }
\bu = (\sqrt{1 - \vars^{2}} \, \cos \theta , \sqrt{1 - \vars^{2}} \, \sin \theta , \vars) ]
\nonumber
\fr
&
= \dfrac{1}{8} \, |\bV| \int_{-1}^{1} d \vars \int_{-\pi}^{\pi} d \theta \:
F(\ve, \vs, \ot (\ve + \vs) + \ot |\bV| \, \bu , 
\ot (\ve + \vs) - \ot |\bV| \, \bu ) \p
\tag{A.3} \label{Fu}
\end{align}
\mbox{ } \\[-15pt]
\emph{The total cross section.} 
\fr
We must consider the integral
\begin{equation}
\ivn{2} W(\ve, \vs | \vp, \vsp) \: d \vp \, d \vsp \p
\tag{A.4} \label{A_nu} 
\end{equation}
We apply the formula \eq{F_A}, where now 
$ F(\ve, \vs, \vp, \vsp) = \Ke(\bn \cdot \bV, |\bV|)$.
\\
Since 
$$ 
\bn = \dfrac{(\ve - \vp)}{|\ve - \vp|} = \dfrac{\bm{A}}{|\bm{A}|} ,
$$
then we have
\begin{equation}
\ivn{2} W(\ve, \vs | \vp, \vsp) \: d \vp \, d \vsp =
\frac{1}{2} \ivn{}  \Ke \left( \dfrac{\bm{A}}{|\bm{A}|} \cdot \bV, |\bV| \right) 
\delta(|\bm{A}|^{2} - \bm{A} \cdot \bV) \: d \bm{A} \p 
\tag{A.5} \label{W_A} 
\end{equation}
If $\bV = \bm{0}$ then the integral \eq{W_A} holds zero, otherwise, we fix the vector $\bV$, 
and we choose a reference frame such that $\bV = (0, 0, |\bV|)$. Then we introduce 
spherical coordinates, so that
$$
\bm{A} = r (\sin \varphi \, \cos \theta , \sin \varphi \, \sin \theta , \cos \varphi) ,
\quad \mbox{which implies} \quad
\dfrac{\bm{A}}{|\bm{A}|} \cdot \bV = |\bV| \cos \varphi \sv
$$
and the integral \eq{W_A} becomes 
\begin{align*}
& 
\frac{1}{2} \int_{0}^{+ \infty}  dr 
\int_{0}^{\pi} \! d \varphi \int_{0}^{2 \pi} \! d \theta \,
\Ke(|\bV| \cos \varphi , |\bV|) \, 
\delta \left( r^{2} - r \, |\bV| \cos \varphi \right) r^{2} \sin \varphi
\\[4pt]
& \mbox{ }
 = \pi \int_{0}^{\pi} \Ke(|\bV| \, \cos \varphi , |\bV|) \left[ 
\int_{0}^{+ \infty} \delta \left( r^{2} - r \, |\bV| \cos \varphi \right) r^{2} \: dr
\right] \sin \varphi \: d \varphi
\\[4pt]
& \mbox{ }
 = \pi \int_{-1}^{1} \Ke(|\bV| \, \vars , |\bV|) \left[ 
\int_{\rea} \delta \left( r^{2} - r \, |\bV| \,\vars \right) r^{2} \, H(r) \: dr
\right] d \vars
\\[4pt]
& \mbox{ } = 
\pi \int_{0}^{1} \Ke(|\bV| \, \vars , |\bV|) \, |\bV| \, \vars \: d \vars \sv
\end{align*}
where $H(r)$ is the Heaviside step function.
Hence the integral \eq{W_A} is a function of $\ve$ and $\vs$ only through the 
modulus of $\bV$, and we can define
\begin{equation}
\nu(|\bV|) :=  \ivn{2} W(\ve, \vs | \vp, \vsp) \: d \vp \, d \vsp = 
\pi \, |\bV| \int_{0}^{1} \Ke(|\bV| \, \vars , |\bV|) \, \vars \: d \vars \p
\tag{A.6} \label{A_freq}
\end{equation}
In the case of \emph{Variable Hard Sphere}, we have
$$
\nu(|\bV|)  = \dfrac{\pi}{2} \, |\bV| \, \mathsf{\Ke}(|\bV|) \p
$$
\emph{The gain kernel $G$.} 
\fr
We study the integral
\begin{equation}
G(\phi; \ve , \vs) = 
\ivn{2} W(\ve, \vs | \vp, \vsp)
\left[ \phi(\vp) + \phi(\vsp) \right] d \vp \, d \vsp
\tag{A.7}
\end{equation}
The apply the formula \eq{Fu}, where 
now 
$ F(\ve, \vs, \vp, \vsp) = \Ke(\bn \cdot \bV, |\bV|) \left[ \phi(\vp) + \phi(\vsp) \right]$.
\\
Since
$$
\ve - \vp = \ve - \ot (\ve + \vs) - \ot |\bV| \, \bu =
\ot (\ve - \vs) - \ot |\bV| \, \bu = \ot \, \bV - \ot \, |\bV| \, \bu \sv
$$
we have
\begin{align*}
\bn \cdot \bV & = \dfrac{(\ve - \vp)}{|\ve - \vp|} \cdot \bV 
= 
\dfrac{ \left( \ot \, \bV - \ot \, |\bV| \, \bu \right) \cdot \bV}{
\left| \, \ot \, \bV - \ot \, |\bV| \, \bu \, \right|}
=
\dfrac{ |\bV |^{2} - |\bV| \, (\bu \cdot \bV) }{ \left| \, \bV - |\bV| \, \bu \, \right|}
\fr
& =
\dfrac{\ot \left| \, \bV - |\bV| \, \bu \, \right|^{2}}{
\left| \, \bV - |\bV| \, \bu \, \right|}
=
\ot \left| \, \bV - |\bV| \, \bu \, \right| .
\end{align*}
Hence
\begin{align}
G(\phi; \ve , \vs) = &
\: \dfrac{|\bV|}{8} \int_{-1}^{1} d \vars \int_{-\pi}^{\pi} d \theta \:
\Ke \left( \ot \, \left| \bV - |\bV| \, \bu \, \right|, |\bV| \right) 
\nonumber
\fr
& \times \left[ \phi \left( \ot (\ve + \vs) + \ot |\bV| \, \bu \right) + 
\phi \left( \ot (\ve + \vs) - \ot |\bV| \, \bu \right) \right] .
\label{Gpu} \tag{A.8}
\end{align}
In the case of \emph{Variable Hard Sphere}, we have
\begin{align}
G(\phi; \ve , \vs) & = 
\dfrac{|\bV|}{8} \, \mathsf{\Ke}(|\bV|)
\int_{-1}^{1} d \vars \int_{-\pi}^{\pi} d \theta \left[
\phi \left( \ot (\ve + \vs) + \ot |\bV| \, \bu \right) +
\phi \left( \ot (\ve + \vs) - \ot |\bV| \, \bu \right) \right]
\nonumber
\fr
& =
\dfrac{|\bV|}{4} \, \mathsf{\Ke}(|\bV|)
\int_{-1}^{1} d \vars \int_{-\pi}^{\pi} d \theta \:
\phi \left( \ot (\ve + \vs) + \ot |\bV| \, \bu \right) .
\tag{A.9}
\end{align}
%

%
%
%
%
%
%

%
\end{document}